\documentclass[twocolumn,amsmath,amssymb,pre,aps,floatfix,a4paper,superscriptaddress,nobibnotes,raggedbottom]{revtex4}
\usepackage{graphicx}
\usepackage[unicode,hidelinks]{hyperref}
\newcommand{\manuscripttitle}{Chaos in active particle memristive neurons}
\newcommand{\manuscriptauthors}{Ying Quan, Fabio Marchesoni, Yunyun Li, Marco G. Mazza, Sergey Savel’ev}
\hypersetup{pdftitle={\manuscripttitle},pdfauthor={\manuscriptauthors}}

\makeatletter
\DeclareRobustCommand{\correspondingmark}{\appdef\@affilID@temp{{*}}}
\long\def\@makecaption#1#2{%
  \par\vspace{2pt}\noindent
  \makebox[\linewidth][c]{\parbox{0.94\linewidth}{\footnotesize\textbf{#1.} #2}}\par}
\newlength{\pairedcolumnwidth}
\newenvironment{modeldisplays}
  {\setlength{\abovedisplayskip}{5pt plus 1pt minus 1pt}%
   \setlength{\belowdisplayskip}{\abovedisplayskip}%
   \setlength{\abovedisplayshortskip}{\abovedisplayskip}%
   \setlength{\belowdisplayshortskip}{\abovedisplayskip}}
  {}
\newenvironment{inplacefigure}
  {\par\addvspace{\intextsep}\noindent
   \begin{minipage}{\columnwidth}\def\@captype{figure}\centering}
  {\end{minipage}\par\addvspace{\intextsep}}
\newenvironment{inlinewidefigure}
  {\onecolumngrid\par\addvspace{8pt}\noindent
   \begin{minipage}{\textwidth}\def\@captype{figure}\centering}
  {\par\end{minipage}\par\addvspace{8pt}\twocolumngrid}
\makeatother

\begin{document}
\title{\manuscripttitle}
\author{Ying Quan}
\author{Fabio Marchesoni}
\author{Yunyun Li\correspondingmark}
\affiliation{MOE Key Laboratory of Advanced Micro-Structured Materials, School of Physics Science and Engineering, Tongji University, Shanghai 200092, China}
\author{Marco G. Mazza}
\affiliation{Interdisciplinary Centre for Mathematical Modelling and Department of Mathematical Sciences, Loughborough University, LE11 3TU, UK}
\author{Sergey Savel’ev}
\affiliation{Department of Physics, Loughborough University, LE11 3TU, United Kingdom}

\begin{abstract}
Inspired by the particle drift and diffusion dynamics in diffusive memristors that resemble those of biological neurons, we design a deterministic active particle memristive neuron model. To form a neuron, memristor is connected with capacitor in parallel and inductor in series. The inductor represents the system inertia, giving rise to a four-dimensional nonlinear dynamical system. Numerical simulations show that the system can exhibit three chaotic transitions: Type-I Pomeau--Manneville intermittency, transient chaos, and chaos raised from period-doubling bifurcations. These results demonstrate that, through bias-voltage tuning, such artificial neurons can operate at the edge of chaos, a regime considered important for neuromorphic computing.
\end{abstract}
\maketitle

\section{Introduction}\label{sec:intro}

Traditional computing follows the von Neumann architecture, in which
data storage and computation are separated. During computation, data
must be frequently read, transferred, and written back between the
CPU/GPU and memory \cite{ref1,ref2,ref3}. As a result, it faces the problems of
high power consumption and limited computing capability when processing
massive data and complex tasks \cite{ref4,ref5}. By contrast, neural-network neural circuits integrate memory and processing within distributed, highly parallel synaptic networks \cite{ref6}, achieving remarkable energy efficiency and adaptive responsiveness \cite{ref7}. These advantages have motivated extensive research in neuromorphic computing and engineering, which seeks to emulate neurobiological computational paradigms at the hardware level \cite{ref8,ref9,ref10}.

Among devices for implementing hardware synapses and neurons, memristors have emerged as prominent building blocks \cite{ref11,ref12,ref13}. Originally proposed by Chua, the memristor was defined as the fourth fundamental passive circuit element \cite{ref14}. Memristor's instantaneous resistance depends on the history of the applied electrical signal, endowing the device with internal-state memory and history-dependent hysteresis \cite{ref15,ref16}. This plasticity mirrors biological synaptic behaviour and enables analogue artificial neuron spiking rate and artificial synapse weight adaptation \cite{ref17}. Physical implementations of memristors encompass a wide range of materials and resistive switching mechanisms \cite{ref18,ref19}. Although these mechanisms can functionally emulate biological synaptic plasticity, their physical origins usually differ from the ionic conduction processes in biological synapses. By contrast, diffusive memristors modulate conductance through the migration and diffusion of metal ions \cite{ref20,ref21}. Their internal state evolution, driven by particle transport, thus shares certain physical similarities with the ion-mediated processes that regulate biological synaptic plasticity \cite{ref22}.

Recently the model in which active particles shuttle between memristive
terminals of a diffusive neuron has been shown to generate diverse spiking modes, including
regular periodic spikes and bursting discharges \cite{ref23}. Owing to the structural and functional asymmetry of
Janus particles, they can exhibit sustained self-propelled motion, thereby introducing an additional timescale into
the system that is distinct from the applied bias signal \cite{ref24,ref25,ref26}. Beyond emulating regular spiking, understanding dynamical transitions between ordered and disordered spiking dynamical states in neural systems is an important task to research in the development of artificial neurons. Previous studies have suggested that certain neural systems may operate near the “edge of chaos” \cite{ref27,ref28}, a regime that may facilitate information propagation, temporal feature extraction, and dynamical computation \cite{ref29,ref30,ref31}. The term “edge of chaos” carries different meanings across theoretical frameworks. Here, we adopt the dynamical systems interpretation, using the term to refer to the vicinity of transitions between periodic and chaotic dynamics, identified through a combination of Lyapunov spectrum and bifurcation analyses.

The circuit proposed here is biologically inspired by the coupling between neuronal membrane potential and state dependent ionic conductance \cite{ref32}. The capacitor represents the membrane capacitance, and the voltage across it corresponds to the membrane potential \cite{ref32}. The active particle memristor connected in parallel with the capacitor serves as a phenomenological model of an effective ion channel, modulating the transmembrane conductance through internal particle motion \cite{ref33}. Previous studies have shown that delayed gating responses of ion channels can impart inductive characteristics to the membrane impedance under certain conditions \cite{ref34}. Motivated by these findings, we introduce an inductor into the circuit to represent an additional timescale of the current response and investigate the dynamics arising from its coupling with the membrane capacitance and the time-varying conductance. We demonstrated that this circuit shows chaotic dynamics and study transition to chaos with increasing applied voltage.

The paper is organized as follows. The active particle memristive neuron model with an inductance is presented in Sec.~\ref{sec:model}. The main numerical results are presented in Sec.~\ref{sec:results}, where we first analyze the bifurcation in the system with respect to the external bias voltage. We then examine three chaotic transitions in turn, namely Type-I intermittency, transient chaos, and period-doubling bifurcation cascade leading to chaos. Finally, conclusions are drawn in Sec.~\ref{sec:conclusions}.
\section{Model}\label{sec:model}

The device is a two-terminal microchannel in which a single conductive,
self-propelled Janus particle moves between two parallel planar electrodes at
$x=x_{\min}$ and $x=x_{\max}$, as sketched in Fig.~\ref{fig:1} (see also
Ref.~\onlinecite{ref23}). The channel is open along the longitudinal
$y$-direction. The Janus particle
generates a localised phoretic flow field under external
excitation, resulting in a persistent self-propulsion speed $v_0$. To
describe the motion of the self-propelled active particle between the
electrodes, we assume an overdamped deterministic dynamic model, namely

\begin{modeldisplays}
\begin{equation}
\frac{dx}{dt} = v_{0}\cos\theta - \frac{\partial U}{\partial x},
\end{equation}
\begin{equation}
\frac{dy}{dt} = v_{0}\sin\theta,
\end{equation}
\begin{equation}
\frac{d\theta}{dt} = \omega.
\end{equation}
\end{modeldisplays}

Here $x$ is the transverse position of the particle centre of mass, $t$ is time, and $\theta$ is the orientation of the self-propulsion velocity relative to the positive $x$-axis; $\omega$ is
instantaneous angular velocity. The coordinate y denotes the longitudinal position of the particle along the channel. We deliberately omit translational and rotational
thermal noise which could hide the deterministic chaotic dynamics. Confinement by the channel sidewalls is modelled by a steeply repulsive potential:

\begin{equation}
U(x)=\varepsilon\left[\left(\frac{\sigma}{x-x_{\min}}\right)^{12}
+\left(\frac{\sigma}{x_{\max}-x}\right)^{12}\right].
\end{equation}

All lengths are normalized by half of the distance between the electrodes, so that the electrodes are located at $x_{\min}=-1$ and $x_{\max}=1$. In the numerical simulations, we set \(\varepsilon = 1.0\ \)and \(\sigma = 0.05\). 
\(U(x)\) is nearly zero in most of the central region and rises
steeply only as the particle approaches either boundary. 

\begin{inplacefigure}
\includegraphics[width=0.98\columnwidth]{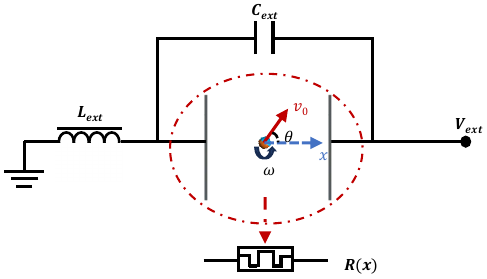}
\caption{Schematic of the memristive neuron circuit model. Here, \(x\) denotes the particle's position, \(v_{0}\) the self-propulsion speed, \(\theta\) the orientation angle, and \(R(x)\) the memristor's resistance. \(C_{ext}\) and \(L_{ext}\) represent the external capacitor and inductor  respectively, and \(V_{ext}\) is the applied bias voltage.}\label{fig:1}
\end{inplacefigure}

Electron transport through the memristor
is modelled as a two-step tunnelling process in an
``electrode--particle--electrode'' configuration. The two tunnelling
resistances depend exponentially on the distances between the particle
and the two electrodes. When the particle is at position \(x\), the
distances are \(d_{1} = 1 - x\) and \(d_{2} = x + 1\). Accordingly,
\(R_{1} \propto e^{(1 - x)/\lambda}\) and
\(R_{2} \propto e^{(x + 1)/\lambda}\), where \(\lambda\) is the
tunnelling length. Under the incoherent tunnelling approximation, the
two resistances are connected in series, and the total resistance
\(R = R_{1} + R_{2}\ \)is determined by the particle position \cite{ref35}:

\begin{equation}
R(x)=R_{0}\cosh\left(\frac{x}{\lambda}\right).
\end{equation}

where $R_{0}$ is the minimal resistance of the memristor, attained when the particle is located at the channel centre. Owing to the strong nonlinearity of \(R(x)\), the device resides in a
low-resistance state when the particle is near the central region,
whereas \(R\) increases sharply as the particle approaches either
electrode, driving the device into a high-resistance state. The applied
bias voltage, together with feedback from the external circuit, can
induce migration of the particle between these regions,
thereby enabling switching between low and high resistance states.

Because the external field exerts an electrokinetic torque on the Janus
particle\cite{ref36,ref24}, its angular velocity $\omega$ depends
nonlinearly on the terminal voltage $V$. Following  Ref.~\onlinecite{ref23} we assume a
sigmoidal activation:

\begin{equation}
\omega=\omega_{0}\tanh\left(\frac{V-V^{*}}{\alpha V^{*}}\right).
\label{eq:omega}
\end{equation} 
where $\omega_0$ is the saturation angular velocity, $V^*$ the voltage
threshold, and $\alpha$ controls the voltage width of angular velocity switching. $\omega$ reverses sign as $V$ crosses $V^*=1.0$ within a transition zone of
width $\alpha V^*$, causing the particle to pivot and reverse its direction of
travel across the gap. Equation~\eqref{eq:omega} also breaks the inversion
symmetry $(x,\theta,I_L,V)\mapsto(-x,\pi-\theta,-I_L,-V)$ that the remaining
equations would otherwise possess, which is the origin of the pronounced bias-polarity asymmetry discussed later.

We consider a neuromorphic circuit ("artificial neuron") made of the active particle memristor in
parallel with an external capacitance $C_{\mathrm{ext}}$ and in series with an external inductance $L_{\mathrm{ext}}$ driven by a
DC bias source $V_{\mathrm{ext}}$. Kirchhoff's laws
give:

\begingroup
\setlength{\abovedisplayskip}{17pt}
\setlength{\belowdisplayskip}{\abovedisplayskip}
\setlength{\abovedisplayshortskip}{\abovedisplayskip}
\setlength{\belowdisplayshortskip}{\abovedisplayskip}
\setlength{\jot}{9pt}
\setlength{\intextsep}{17pt}
\begin{equation}
\frac{dI_{L}}{dt}=\frac{V_{ext}-V}{L_{ext}},
\end{equation}
\begin{equation}
\frac{dV}{dt}=\frac{I_{L}-V/R(x)}{C_{ext}}.
\end{equation}

for the series inductor current $I_L$ and the memristor terminal voltage $V$. The
current through the memristive branch, which we use throughout as the 
observable, is $I_R(t)=V(t)/R\big(x(t)\big)$.

By combining Eqs (1) -- (8), we thus obtain a four-dimensional
dynamical system of the form:

\begin{equation}
\left\{
\begin{aligned}
\dot{x} &= v_{0}\cos\theta
+12\varepsilon\sigma^{12}
\left[\frac{1}{(x+1)^{13}}-\frac{1}{(1-x)^{13}}\right],\\
\dot{\theta} &= \omega_{0}\tanh\left(\frac{V-V^{*}}{\alpha V^{*}}\right),\\
\dot{I}_{L} &= \frac{V_{ext}-V}{L_{ext}},\\
\dot{V} &= \frac{1}{C_{ext}}\left(I_{L}-\frac{V}{R_{0}\cosh(x/\lambda)}\right).
\end{aligned}
\right.
\end{equation}

In this work, we use the fourth-order Runge--Kutta scheme to solve the
system of ordinary differential equations given in Eq (9). Unless
otherwise stated, the parameters are set to \(R_{0} = 1.0\),
\(\lambda = 0.12\), \(V^{*} = 1.0\), \(\alpha = 1.0\),
\(\omega_{0} = 1.0\), \(v_{0} = 1.0\), \(L_{ext} = 20\),
\(C_{ext} = 0.005\). Some of the parameters are taken from Ref.~\onlinecite{ref23}, which pointed out that the values of the Janus particle parameters are consistent with experimental observations when lengths are measured in micrometres and times in seconds. In the same units, as $R(x)$ varies between the low and high resistance states, the system time scales $RC_{ext}$ and $L_{ext}/R$ range from milliseconds to tens of seconds, thus spanning the characteristic time scale of the particle motion. This enables an effective feedback between the circuit dynamics and the particle motion. The initial conditions $(x_{0},\theta_{0},I_{L,0},V_{0})=(0.1,0,0,0)$ are used in all simulations.

\newpage
\section{Results and Discussion}\label{sec:results}

We first analyse the equilibrium points of this system. Setting
$\dot{I}_L=0$ requires $V=V_{\mathrm{ext}}$, whereas $\dot{\theta}=0$ requires
$V=V^{*}$. These two conditions are simultaneously satisfiable only if

\begin{equation}
	V_{\mathrm{ext}} = V^{*} .
	\label{eq:eqcondition}
\end{equation}

Consequently, for every $V_{\mathrm{ext}}\neq V^{*}$ the system has no
	equilibrium anywhere in phase space. Since the flow is dissipative
 and, as we verify numerically, all trajectories remain
bounded, the asymptotic state
must be a nonstationary invariant set: periodic, quasiperiodic, or chaotic.
Self-sustained firing is therefore structurally guaranteed rather than the result
of a Hopf bifurcation.

\begin{inplacefigure}
\includegraphics[width=1.00\columnwidth]{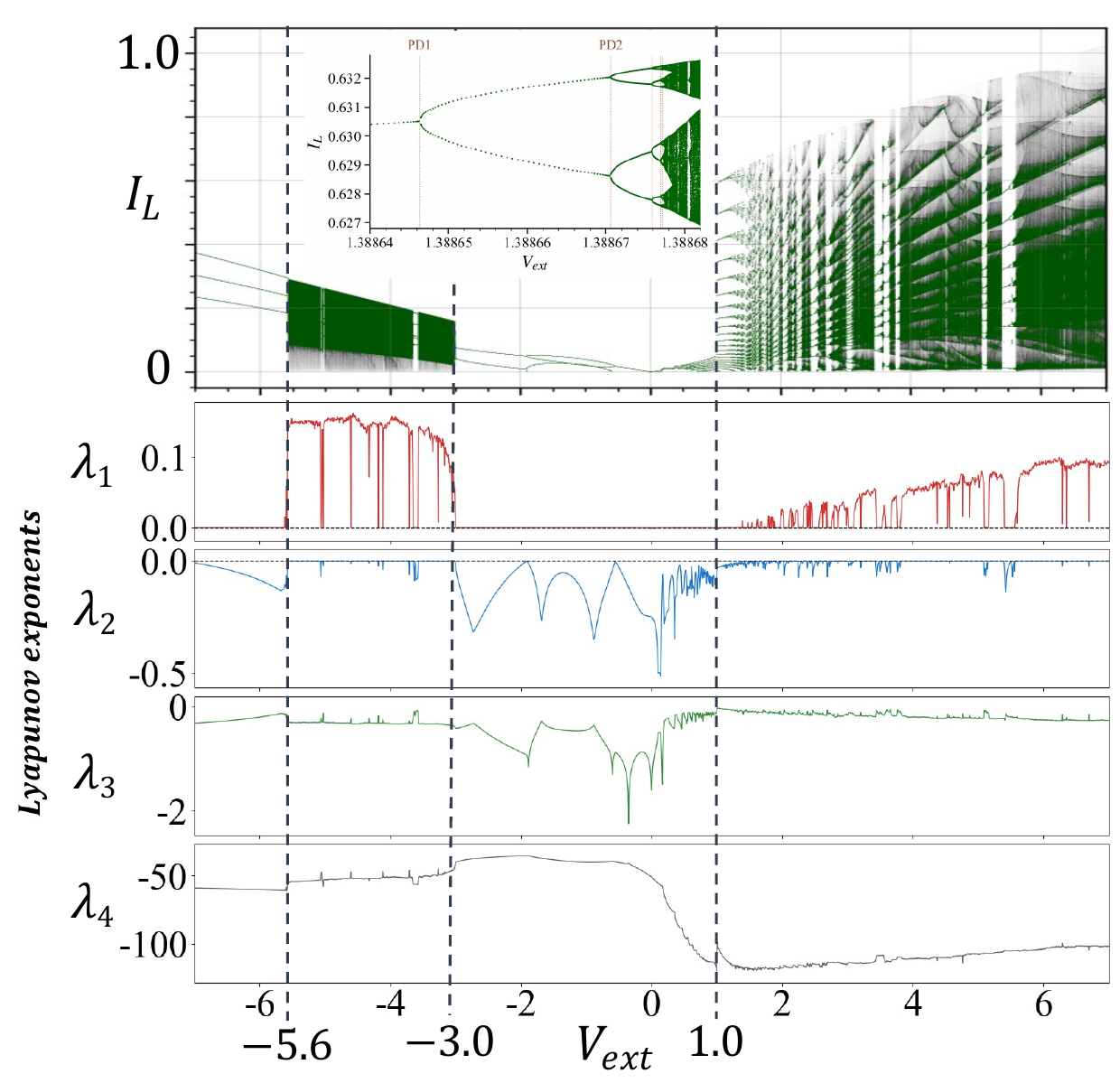}
\caption{Bifurcation diagram of \(I_{L}\) recorded on the Poincar\'e section \(V=0\) with \(\mathrm{d}V/\mathrm{d}t>0\) versus \(V_{ext}\) (top) and the corresponding Lyapunov exponents \(\lambda_{1}\)--\(\lambda_{4}\) (bottom). The inset magnifies the period-doubling bifurcation near \(V_{ext}\approx1.3886\).}\label{fig:2}
\end{inplacefigure}
\endgroup

Fig.~\ref{fig:2} shows the bifurcation diagram and the corresponding Lyapunov exponent spectrum, revealing the effect of the external voltage $V_{ext}$ on the system dynamics. The system is chaotic approximately in the intervals $V_{ext}\in(-5.6,\,-3.0)$ and $(1.39,\,7.0)$, where the largest Lyapunov exponent $\lambda_{1}$ is positive and the bifurcation diagram consists of densely scattered points. Several periodic windows, in which $\lambda_{1}$ drops to zero, are embedded within these two chaotic intervals.

\setlength{\pairedcolumnwidth}{\columnwidth}
\clearpage
\onecolumngrid
\noindent\begin{minipage}[t]{\pairedcolumnwidth}
\vspace{0pt}\setlength{\intextsep}{0pt}
\begin{inplacefigure}
\includegraphics[trim=22bp 0bp 72bp 25bp,clip,width=1.00\columnwidth]{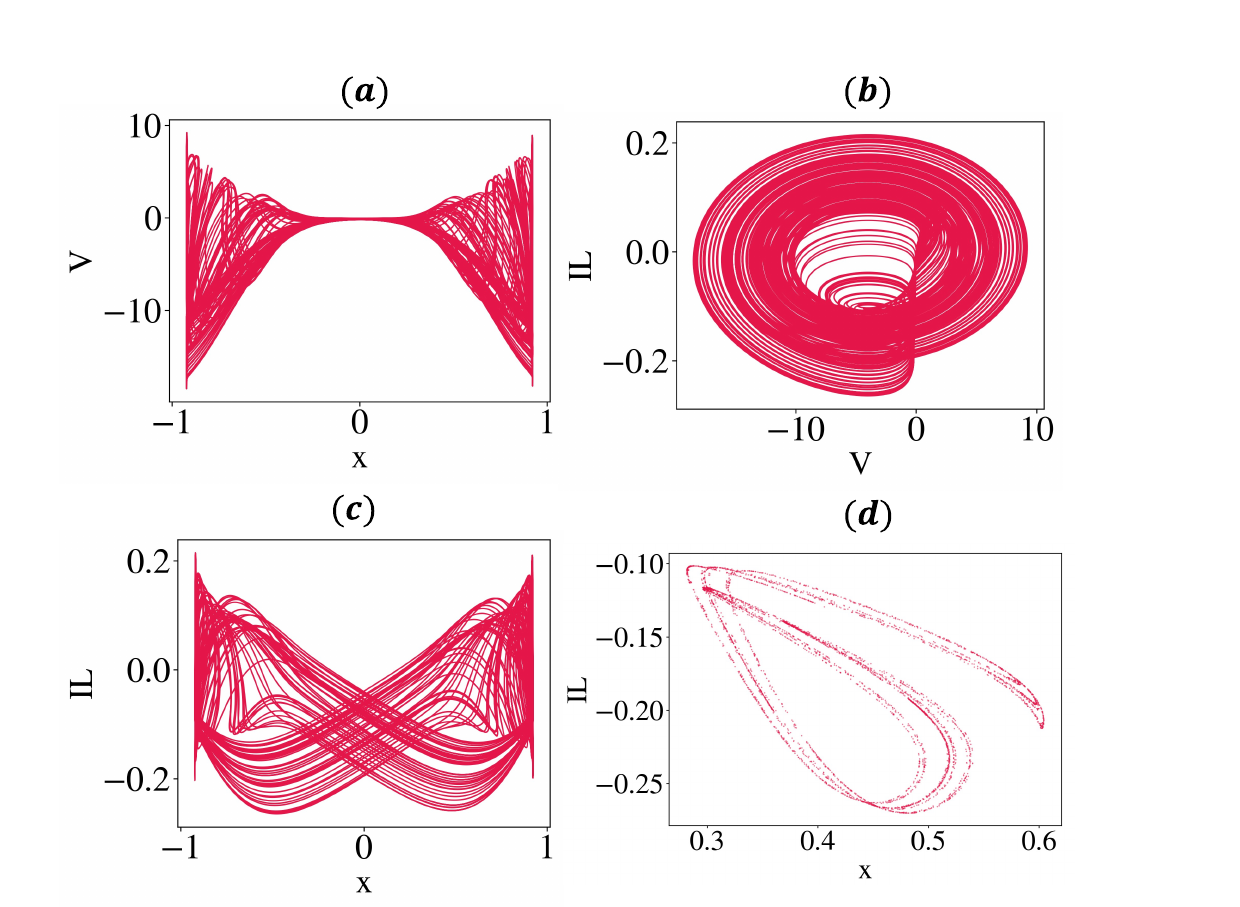}
\caption{Projections of the chaotic attractor for \(V_{ext}=-4\) onto (a) the \(x\)--\(V\) plane, (b) the \(V\)--\(I_{L}\) plane and (c) the \(x\)--\(I_{L}\) plane; (d) projection of the Poincar\'e section \(\theta=0\ (\mathrm{mod}\ 2\pi)\) with \(\mathrm{d}\theta/\mathrm{d}t>0\) of the attractor onto the \(x\)--\(I_{L}\) plane.}\label{fig:3}
\end{inplacefigure}
\end{minipage}\hfill
\begin{minipage}[t]{\pairedcolumnwidth}
\vspace{0pt}\setlength{\parindent}{1em}\indent
Fig.~\ref{fig:3} shows the projections of the chaotic attractor for $V_{\mathrm{ext}}=-4$
onto several two-dimensional planes, together with its Poincar\'e section defined
by $\theta=0\ (\mathrm{mod}\ 2\pi)$ with $\dot{\theta}>0$. The section points lie
on layered and folded bands, which is characteristic of chaos.

Near the critical external bias voltage \(V_{c} = - 3.0049\), the system
dynamics is highly sensitive to the control parameter \(V_{ext}\). When
\(V_{ext} \geq V_{c}\), the trajectory converges to a periodic
orbit, whereas when \(V_{ext} < V_{c}\), the system exhibits chaotic
responses. Numerical simulations show that as \(V_{ext}\) is decreased
from \(- 3.0049\) to \(- 3.2\), the system evolves from periodic
oscillations through intermittent chaos and finally enters fully
developed chaos. As shown in Fig. \ref{fig:4}, intermittent chaos in the
\(V\)--\(t\) time series is characterized by alternate sequences of
long duration, near-periodic laminar phases and short duration, large
amplitude irregular chaos phases. Moreover, as \(V_{ext}\) is
decreased, chaos events occur more frequently and the mean laminar
duration time becomes shorter.
\end{minipage}\par

\begin{inlinewidefigure}
\includegraphics[width=0.94\textwidth]{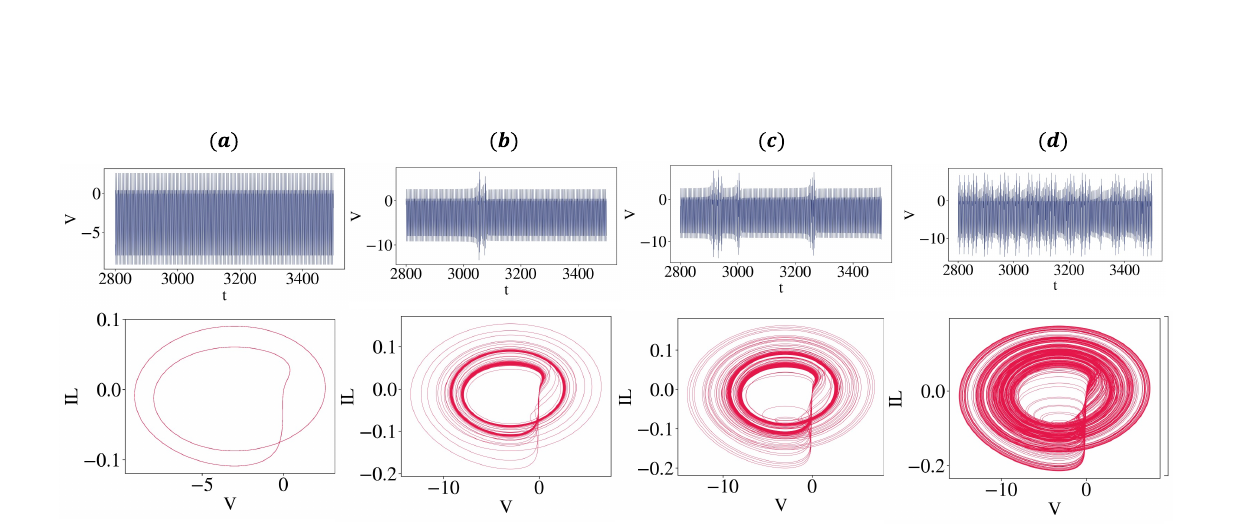}
\caption{\(V\)--\(t\) and \(V\)--\(I_{L}\) phase portraits under different bias voltages \(V_{ext}\). (a) \(V_{ext}=-3.0049\); (b) \(V_{ext}=-3.0050\); (c) \(V_{ext}=-3.01\); (d) \(V_{ext}=-3.2\).}\label{fig:4}
\end{inlinewidefigure}

To quantitatively characterize the intermittent behavior in the periodic--chaotic transition near the critical point, we adopt a criterion based on a Poincar\'e section and the associated return map to distinguish laminar phases from chaos phases~\cite{ref37}. The Poincar\'e section is taken at $V=1$ with $\mathrm{d}V/\mathrm{d}t>0$. Denoting by $I_L(n)$ the inductor current at the $n$th crossing of this section by the trajectory $(V(t), I_L(t), x(t), y(t), \theta(t))$, we construct the return map $I_L(n)\rightarrow I_L(n+1)$ (Fig.~\ref{fig:5}). At $V_{ext}=V_c$, the periodic orbit intersects the section at a single point lying on the diagonal $I_L(n+1)=I_L(n)$, with $I_L\approx 0.064$. If two successive section points satisfy $I_L(n),\,I_L(n+1)\in(0.060,0.070)$, and at least four consecutive section points meet this condition, the time interval between the first and the last of these crossings is identified as a laminar phase.

\onecolumngrid
\clearpage
\begin{inlinewidefigure}
\includegraphics[width=0.79\textwidth]{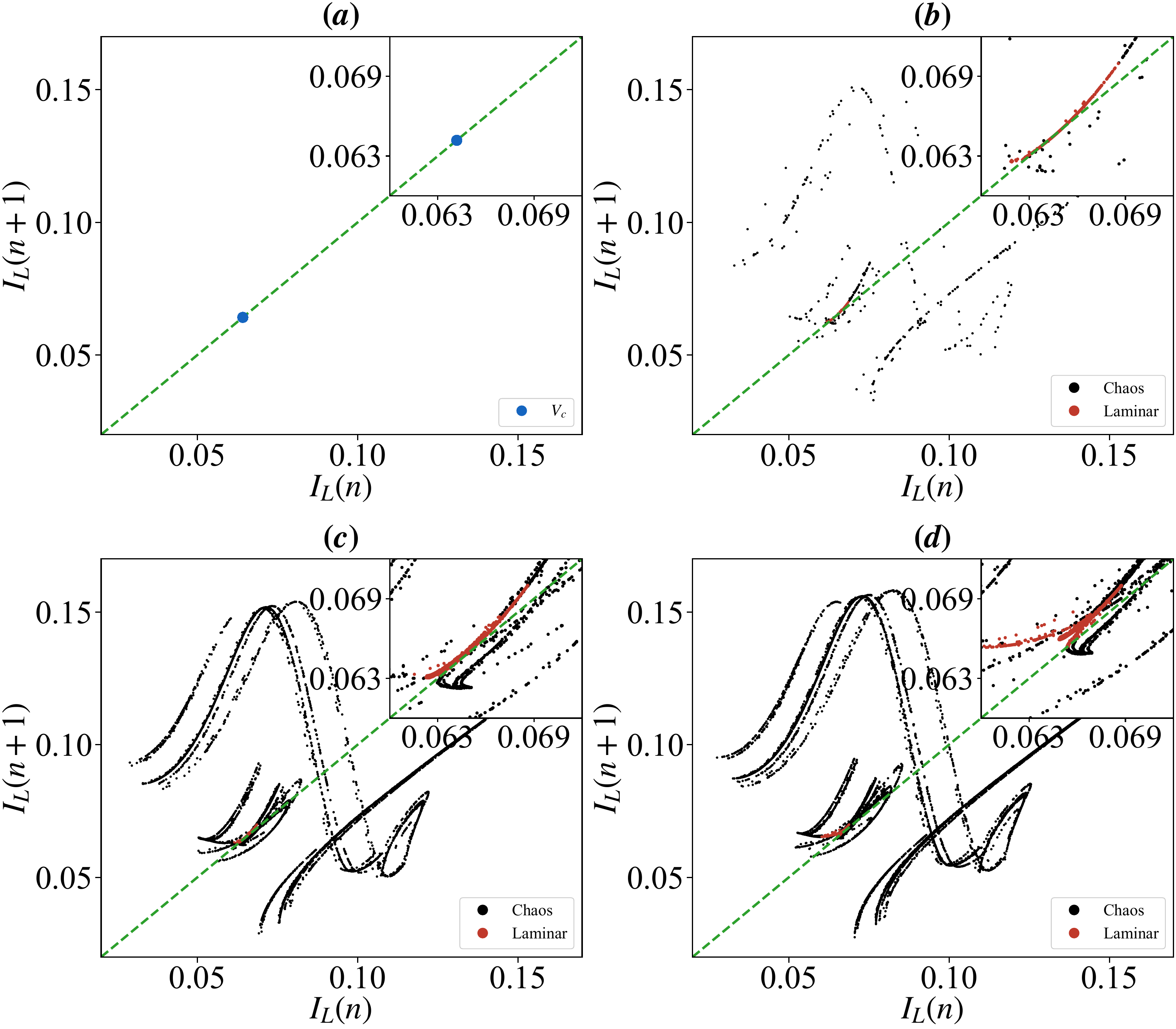}
\caption{Return maps \(I_{L}(n)\rightarrow I_{L}(n+1)\) on the Poincar\'e section \(V=1\) with \(\mathrm{d}V/\mathrm{d}t>0\) under different external bias voltages: (a) \(V_{ext}=-3.0049\), (b) \(-3.0050\), (c) \(-3.02\) and (d) \(-3.1\). Blue dots denote the periodic orbit, black dots denote the chaos phase and red dots denote the laminar phase; the dashed line is the diagonal \(I_{L}(n+1)=I_{L}(n)\). Insets show magnified views of the region near the diagonal.}\label{fig:5}
\end{inlinewidefigure}

As shown in Fig.~\ref{fig:5}, a narrow channel is left between the laminar points and the diagonal. Once the trajectory enters the channel, the increment between successive iterates is very small and $I_L(n)$ increases slowly and monotonically along the channel. After leaving the channel, the iterates rapidly move away from the diagonal into a more dispersed black point cloud, corresponding to the chaos phase. Moreover, as $V_{ext}$ moves away from $V_c$, the points corresponding to the laminar intervals gradually move away from the diagonal, and the degree of chaos is intensified. Figure~\ref{fig:6} shows the $V$--$t$ waveform and the $V$--$I_L$--$x$ phase portrait for $V_{ext}=-3.02$. In Fig.~\ref{fig:6}(b), the blue trajectory corresponds to the periodic orbit at $V_{ext}=V_c$. During the laminar phases the trajectory mainly winds around this periodic orbit, whereas during the chaos phases it deviates significantly from this periodic backbone and expands into a larger region of phase space. In Fig.~\ref{fig:6}(c), we scan 300 equally spaced values of $V_{ext}$ between $-3.01$ and $-3.0051$, where $\epsilon=|V_{ext}-V_c|$ measures the distance to the critical value $V_c$ and $\langle\tau\rangle$ is the mean laminar duration obtained statistically for each parameter value. The blue points are distributed approximately along a straight line, and the red dashed fit gives a slope of $\gamma\approx-0.5089$, close to the theoretical scaling law of type-I intermittency, $\langle\tau\rangle\propto\epsilon^{-1/2}$ as $\epsilon\rightarrow0^{+}$~\cite{ref37,ref38}.

\begin{inlinewidefigure}
\includegraphics[width=0.84\textwidth]{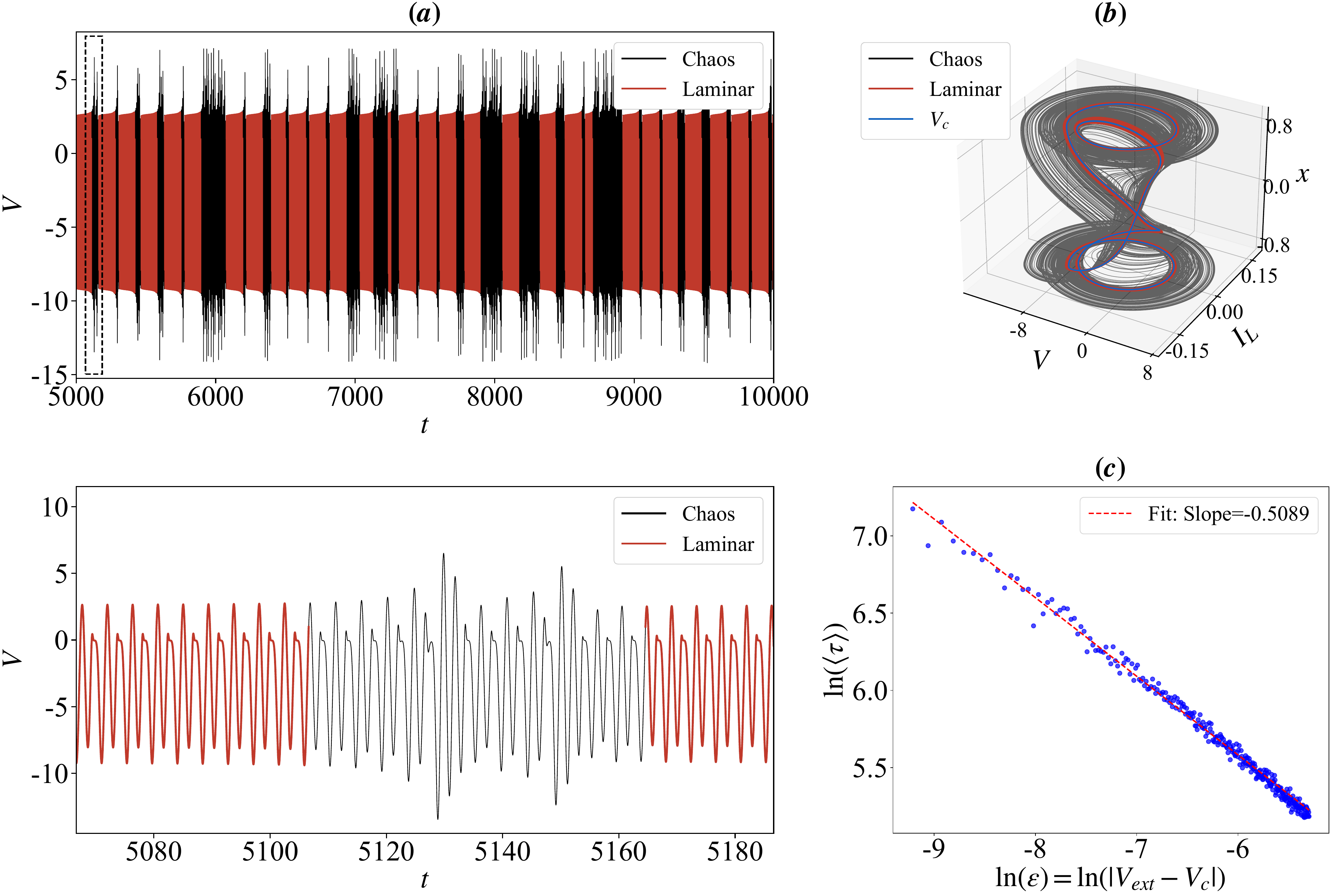}
\caption{In (a) and (b), \(V_{ext}=-3.02\); red denotes the laminar phases and black denotes the chaos phases. (a) \(V\)--\(t\) waveform; the lower panel enlarges the interval marked by the dashed box. (b) \(V\)--\(I_{L}\)--\(x\) phase portrait; the blue trajectory is the periodic orbit at \(V_{ext}=V_{c}\). (c) \(\ln\langle\tau\rangle\) versus \(\ln\epsilon\) for 300 equally spaced values of \(V_{ext}\in[-3.01,-3.0051]\), where \(\langle\tau\rangle\) is the mean laminar duration time; blue dots are simulation data and the red dashed line is a linear fit with slope \(\gamma\approx-0.5089\).}\label{fig:6}
\end{inlinewidefigure}

The system exhibits transient chaos near $V_{ext}=-5.5$. At this external voltage, a chaotic attractor coexists with a periodic attractor, as shown in Fig.~\ref{fig:7}(a). Starting from the initial condition $(x_0,\theta_0,V_0,I_{L0})=(0.1,0,0,0)$, the system remains chaotic throughout the simulation (purple). In contrast, starting from $(-0.491,3.076,-4.577,0.065)$, it converges to a stable periodic oscillation (green). For $V_{ext}=-5.565$ [Fig.~\ref{fig:7}(b)], the system first exhibits chaotic dynamics over a long simulation time (purple) and then converges to a periodic oscillatory state (green). For $V_{ext}=-5.6$ [Fig.~\ref{fig:7}(c)], the system rapidly approaches a regular periodic state.

\begin{inlinewidefigure}
\includegraphics[width=0.98\textwidth]{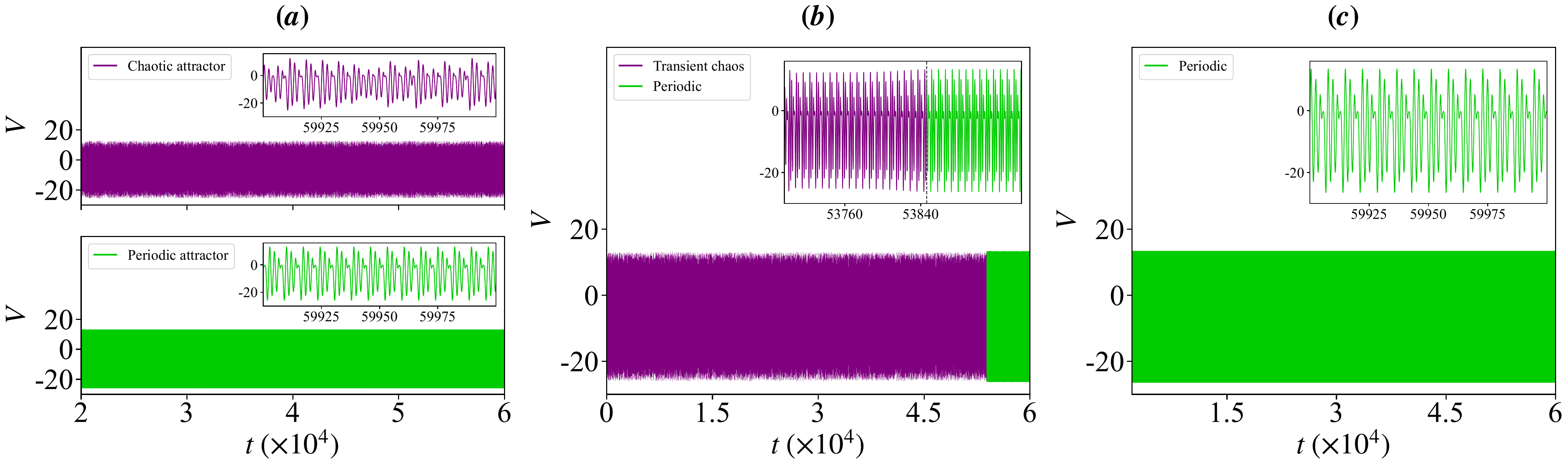}
\caption{Transient chaos: \(V\)--\(t\) time series for (a) \(V_{ext}=-5.5\), (b) \(V_{ext}=-5.565\), and (c) \(V_{ext}=-5.6\). In (a), the chaotic attractor obtained from \((x_{0},\theta_{0},V_{0},I_{L,0})=(0.1,0,0,0)\) coexists with a periodic attractor obtained from \((x_{0},\theta_{0},V_{0},I_{L,0})=(-0.491,3.076,-4.577,0.065)\). Purple indicates the chaotic oscillation state, and green indicates the periodic oscillation state.}\label{fig:7}
\end{inlinewidefigure}

To further elucidate the mechanism of transient chaos in the memristive neuron near $V_{ext}\approx-5.56$, we take the Poincar\'e section $I_L=0$ with $\mathrm{d}I_L/\mathrm{d}t>0$ and construct the return map $V(n)\rightarrow V(n+1)$ (Fig.~\ref{fig:8}). At $V_{ext}=-5.5$ (Fig.~\ref{fig:8}(a)), the black and green points correspond to the coexisting chaotic and periodic attractors, respectively. At $V_{ext}=-5.565$ (Fig.~\ref{fig:8}(b)), the original chaotic attractor is destroyed, and an escape channel opens between the tip of the chaotic set and the periodic point (red arrow). The trajectory first wanders near the remnant chaotic saddle for a long time, corresponding to transient chaos in the time series; after escaping through this channel, the system converges to a stable periodic orbit (green points).

\begin{inlinewidefigure}
\includegraphics[width=0.68\textwidth]{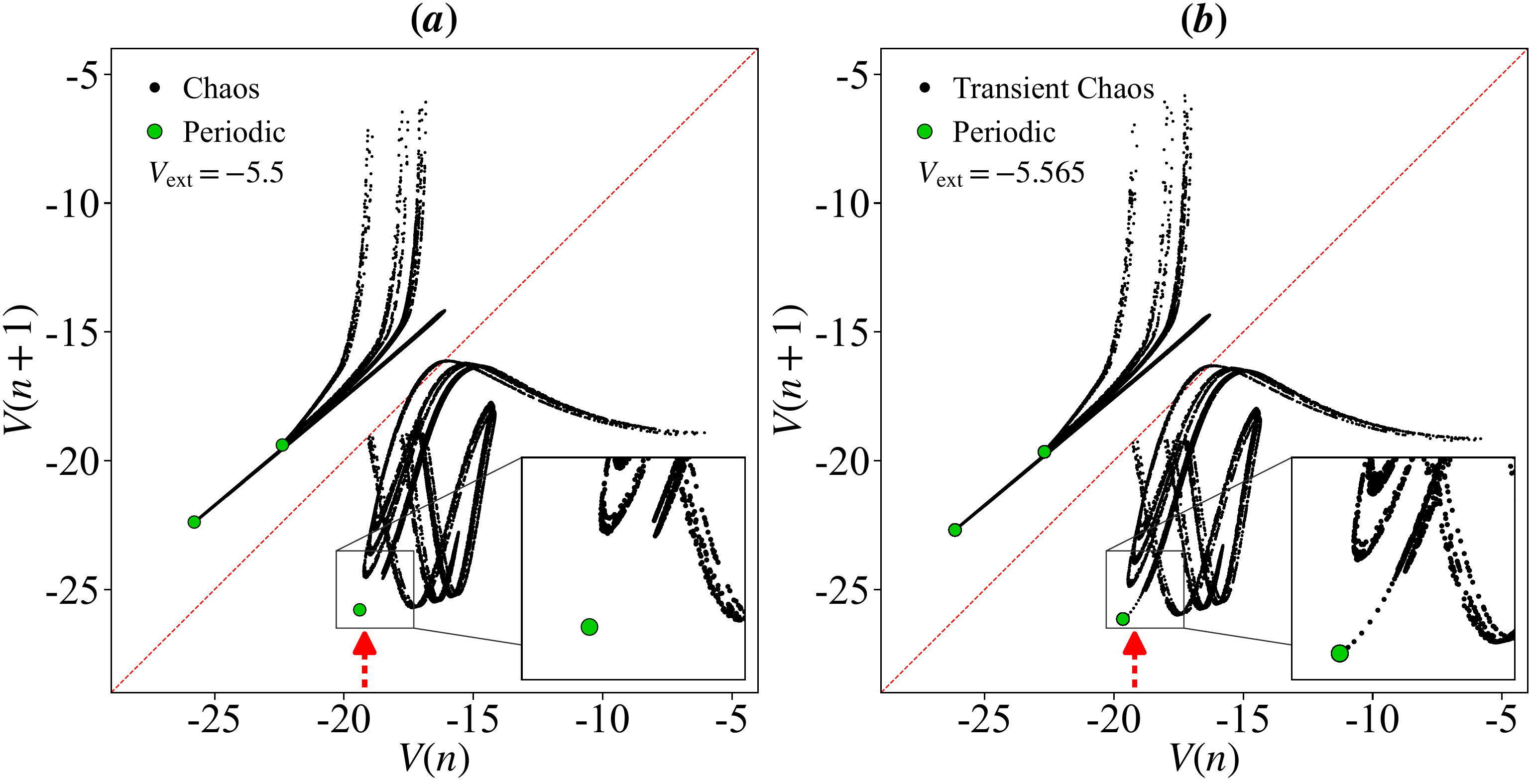}
\caption{ Return maps \(V(n)\rightarrow V(n+1)\) on the section \(I_{L}=0\) with \(\mathrm{d}I_{L}/\mathrm{d}t>0\). (a) Coexisting attractors at \(V_{ext}=-5.5\). (b) At \(V_{ext}=-5.565\), an escape channel forms between the transient chaos and the periodic orbit (marked by the arrows).}\label{fig:8}
\end{inlinewidefigure}

For $V_{ext}>1$, the system enters chaos through period-doubling bifurcations cascade. Fig.~\ref{fig:9} shows the $V$--$I_L$ phase portraits and the corresponding return maps at four values of $V_{ext}$. The insets and the return maps show that the number of intersections of the trajectory with $V=0$ and the number of points in the return maps change from one to two and then to four, and the period continues to double with increasing $V_{ext}$, until finally the intersections spread into a band as the system becomes chaotic.

\begin{inlinewidefigure}
\includegraphics[width=0.98\textwidth]{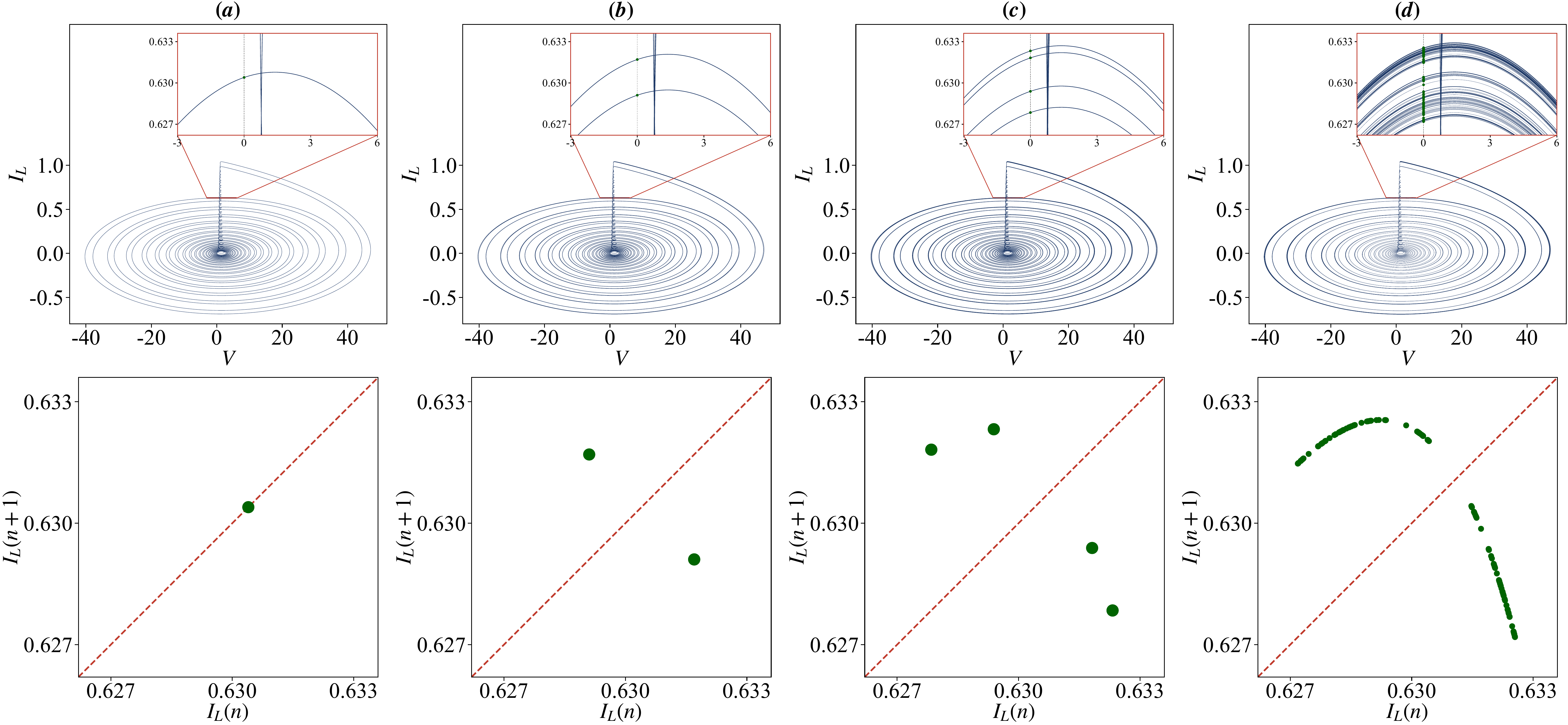}
\caption{\(V\)-\(I_{L}\) phase portraits (upper row) and the corresponding return maps \(I_{L}(n)\text-I_{L}(n+1)\) (lower row) for (a) \(V_{ext}=1.3886402\) (period-1), (b) \(1.3886602\) (period-2), (c) \(1.3886750\) (period-4) and (d) \(1.3886800\) (chaos), showing the period-doubling bifurcation. The return maps are taken on the Poincare section \(V=0\) with \(\mathrm{d}V/\mathrm{d}t>0\), using only the branch with the largest \(I_{L}\).}\label{fig:9}
\end{inlinewidefigure}

\section{Conclusions}\label{sec:conclusions}

We investigate  an artificial  neuron based on active matter  memristor in parallel with capacitance and in series with inductance. It demonstrate diverse spiking regimes when driven by the applied DC external voltage. The artificial neuron displays three chaotic transitions: Type-I Pomeau--Manneville intermittency, transient chaos, and chaos raised from period-doubling bifurcations.

Near the critical bias voltage \(V_{c}\), the system
exhibits typical intermittent chaos, where long-duration, near-periodic
laminar phases alternate with short, irregular chaos phases. By
statistically analyzing the laminar phases, we concluded that the mean
laminar duration \(\langle\tau\rangle\) satisfies a power-law scaling
with the distance \(\epsilon\) between the bias voltage and the critical
point. This scaling law
establishes a predictable quantitative mapping between how far the input
bias is from the critical point and observable temporal statistics,
thereby enabling highly sensitive weak-signal detection.

Moreover, we reveal a transient chaos phenomenon in another parameter
range, whereby the system can exhibit chaotic oscillations for a finite
time and then escape and converge to a stable periodic attractor,
indicating that chaos is not necessarily the system's final attractor.
Further return-map analysis shows that the transient chaos. These results imply that, by adjusting \(V_{ext}\), one can flexibly
balance rich transient chaotic features against stable periodic outputs,
thereby providing a tunable physical dynamical computing primitive for
neuromorphic sensing and computation.

\end{document}